**An EEG-based Stereoscopic Research to Reveal the Brain's Response to What Happens Before and After Watching 2D and 3D Movies**

*Corresponding Author:*
**Negin MANSHOURI**
Karadeniz Technical University
Faculty of Engineering
Department of Electrical and Electronics Engineering
61080 Trabzon/TURKEY
Phone (mobile): (+90) 5535791640
E-mail: n_manshoori90@yahoo.com

**Masoud MALEKI**
Gumushane University
Faculty of Engineering
Department of Electrical and Electronics Engineering
Gumushane/TURKEY
E-mail: mesutmeleki@gumushane@edu.tr

**Temel KAYIKCIOGLU**
Karadeniz Technical University
Faculty of Engineering
Department of Electrical and Electronics Engineering
61080 Trabzon/TURKEY
E-mail: tkayikci@ktu.edu.tr

**An EEG-based Stereoscopic Research to Reveal the Brain's Response to What Happens Before and After Watching 2D and 3D Movies**

**Negin MANSHOURI, Masoud MALEKI , Temel KAYIKCIOGLU**

---

***Abstract***

Despite knowing the reality of three-dimensional (3D) technology in the form of eye fatigue, this technology continues to be retained by people (especially the young community). To check what happens before and after watching a 2D and 3D movie and how this condition influences the human brain's power spectrum density (PSD), a five-member test group was arranged. In this study, electroencephalogram (EEG) was used as a neuroimaging method. EEG recordings of five individuals were taken both before and after watching 2D and 3D movies. After 2D/3D EEG recording, this record was divided into three stages for analysis. These stages consisted of Relax, Watching, and Rest. This benchmarking analysis included I) before and after watching the 2D movie (R2b and R2a), II) before and after watching the 3D movie (R3b and R3a), and III) after watching the 2D/3D movie (R2a and R3a). In the Relax and Rest stages, the 2D/3D EEG power differences in all channels of brain regions for the five EEG bands, including delta (δ), theta (θ), alpha (α), beta (β), and gamma (γ), were analyzed and compared. The PSD based on short-time Fourier transform (STFT) was used to select the dominant bands in this study. Feature extraction was performed on the preprocessed EEG signals using STFT and discrete wavelet transform (DWT). In the 2D analysis, δ, θ, α, and β acted as dominant bands, in 3D, δ, θ, and α were dominant bands, and in 2D/3D, δ and α were selected as meaningful and dominant bands. Partial least-squares regression (PLSR) and support vector machine (SVM) classification algorithms were considered in order to classify the obtained R2a and R3a EEG signals. After the dominant band selection, the correct choice of effective channel combinations in these bands resulted in the percentage of high success of classifiers for Stage III.

## 1. *Introduction*

The three-dimensional (3D) technology has a very long history. Over time, the 3D invention enabled the construction and monitoring of 3D videos, films, and TV programs. The impression of this technology is based on the emergence of photography art in the 18th century. To take 3D photographic images, the stereoscope was invented in the mid-18th century. Then, a stereo animation camera known as Kinetoscope was invented. The development of this technology since the 18th century is illustrated in Fig. 1 [1].

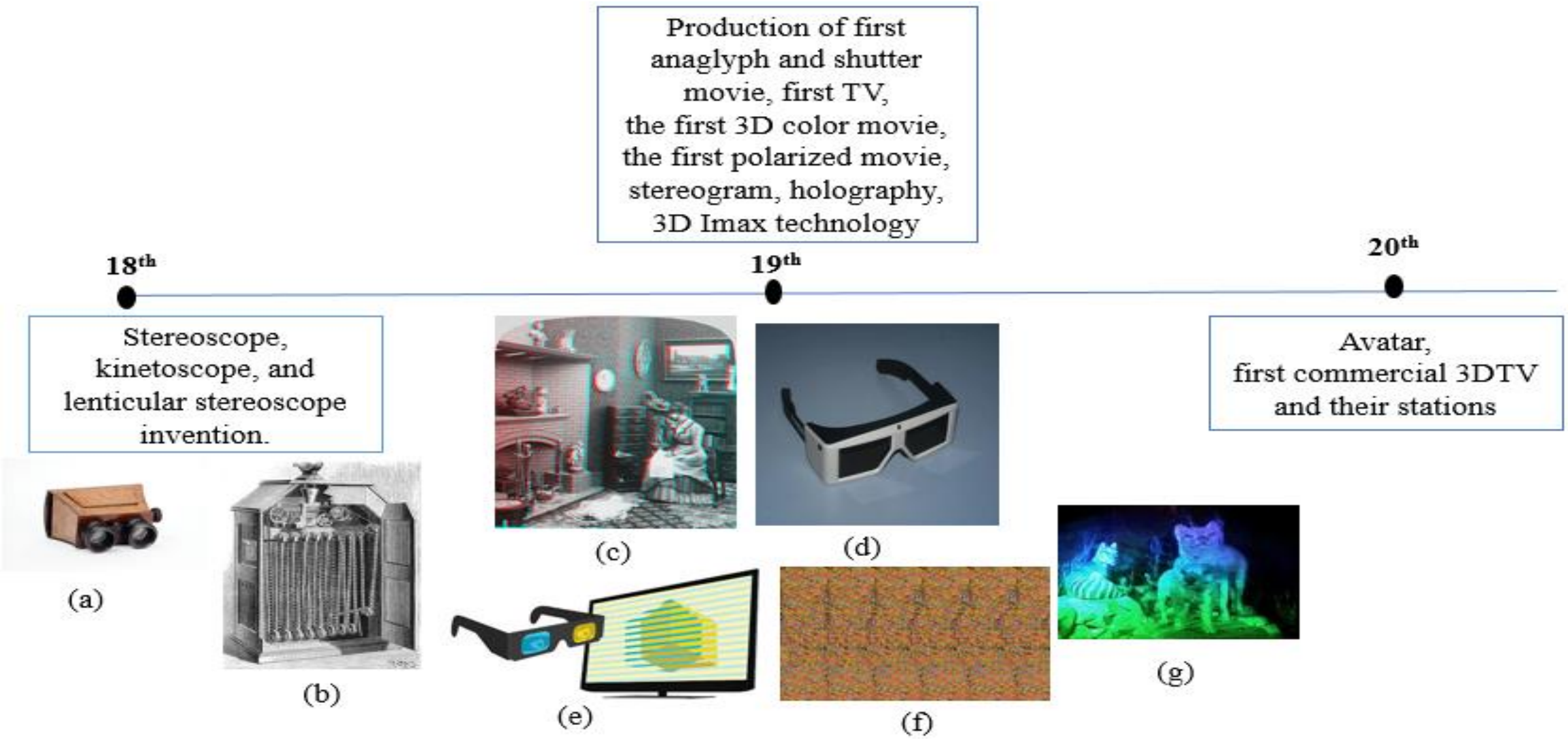

Fig.1. Development of 3D technology since the 18th century, (a) Stereoscope, (b) Kinetoscope, (c) Anaglyph, (d) Shutter, (e) Polarized, (f) Stereogram, (g) Holography [All photos are taken from Wikipedia.]

Thanks to their stereoscopic vision, humans and predators perceive their environment with depth perception. In this vision, each eye sees similar images but with a slight angle difference. This difference is because of the side-by-side location of eyes with an approximately 6.3cm distance. The perception of the depth and correct judgment of the distance are the important achievements of stereoscopic vision [2]. There are different cues such as accommodation, occlusion, motion parallax, and perspective helping the human visual monocular system to perceive depth [3]. The benefits of these cues are a relative perception of depth, even with one eye. 2D TVs are traditional systems using these cues.

There are numerous applications for 3D display, including remote control, scientific and medical visualization, 3D visualization screen design [4], computer design, and entertainment. The application of 3D technology in entertainment has attracted considerable attention from users due to the excitement created [5]. Gaming, console, and animation designer companies greatly benefit from 3D technology. In the market, the 3D product promotion has attracted great attention in terms of advertising. In addition to the expansive use of 3D technology, some disadvantages include cost, health problems (especially eye fatigue), physical discomfort, stress, and habit [6] but because of its realistic and entertaining features, 3D movies are preferred to 2D ones.

There are three kinds of glasses common in this technology: anaglyph, passive, and active. The advantages, disadvantages, and operating principle of these glasses are presented in Table 1.

Table 1. Advantages and disadvantages of 2D and 3D glasses

| Feature | Anaglyph | Active shutter | Passive polarized |
|---|---|---|---|
| How it works | This glass by using two colored filters (blue/cyan and red channels) prepares right and left eye views in different colors [7]. | Every lens of this glass by turns produces images for the right and left eyes at very high speed [8] | In order to create the perception of depth, it prepares the eyes by applying two different polarized images simultaneously [8] |
| Brightness | Depending on the color used in the filter | It has a moderate brightness as it causes a measurable decrease in light transmission. | Thanks to the polarized filters used in these glasses, they have a good brightness due to the minimum effect of light transmission. |
| Availability | High | High, for all TV technology including projection, plasma, LCD and LED LCD | Average, only for LCD, LED LCD |
| Compatibility | High | Low | High, these cinema-type glasses are compatible with all TV sets. |
| Ease of Use | Only require wearing | Due to its connection and synchronization, it has medium ease of use. | It is easy to use without any connection or synchronization. |
| Cost | Inexpensive<br>(Because of the simple electrical nature) | Expensive<br>(In terms of the need for a battery and wireless connection) | Inexpensive<br>(Because of the simple electrical nature) |
| Weight | Generally super lightweight | It is heavy due to the battery usage. | Thanks to its thin plastic lenses and frame it is lightweight. |
| Resolution | Low resolution, relatively poor 3D image quality [9] | It provides good image quality with full resolution. But because of the rapid opening and closing of the shutter mechanism, flickering may lead to head pain negativity. | This passive system with half-vertical resolution sometimes causes black lines on the screen. This problem can be reduced by increasing the distance to the screen or using a smaller screen. |

Neurons of the brain have a special status in the control of actions and behaviors caused by different stimuli in humans. In short, the neuron is known as carrying messages between the human body and the brain. The obtained brain signals or images are used to predict the cognitive behavior of the human brain to various stimuli. Each human behavior has a certain frequency range, greatly improving the understanding of the complex structure of the brain [10].

After an electrical impulse, the nerve cells quickly communicate with each other, resulting in electroencephalogram (EEG) signals. In fact, this signal is the result of cortical nerve cell inhibition and stimulatory postsynaptic potentials [11]. This signal is measured by placing electrodes on different parts of the scalp. According to the frequency ranges of the EEG signals from 0.1 to 100 Hz, these signals are divided into five subdivisions of delta (δ, 0.1-4), theta (θ, 4-8), alpha (α, 8-13), beta (β, 13-25), and gamma (γ, >25). The amplitude of the non-invasive EEG signal is around 100 µV.

In spite of the non-stationary, nonlinear, and non-Gaussian nature of EEG, these signals play an undeniable role in detecting diseases. From among different neuroimaging methods such as transcranial magnetic stimulation (TMS), computed tomography (CT), magnetic resonance imaging (MRI), functional magnetic resonance imaging (fMRI), and EEG [12], EEG is the most preferred technique by researchers due to its ease of use and portability [13].

Information from patterns of EEG signals provides a preliminary view of disease diagnosis. In the early diagnosis of Alzheimer's disease [14], [15], epilepsy [16], [17], tumors [18], [19], dementia [20], [21], brain injury [22], [23], and many other neurological diseases, suitable EEG signal processing facilitates and accelerates the diagnosis of the problem by doctors and researchers. Using an appropriate signal processing method and applying an effective classifier, EEG abnormal signals can be easily distinguished from normal signals.

In addition to biomedical and neurological fields, EEG signals are included in many studies in order to satisfy researchers' curiosity. These signals are the best choice, offering useful information in neuro-marketing [24], game analysis [25], virtual object control [26], social interaction [27], brain-computer interfaces (BCI), and in some research studies, e.g. 2D and 3D gaming effect on the brain signal [8] and investigating visual fatigue of 3D TV [28].

Despite the complexity of 3D technology, some studies have reported that performance is improved in terms of model visualization in comparison with 2D technology [29].

By using EEG-based 2D and 3D concentration games [13], a system of neurofeedback which may be effective in medical areas has been developed. Because of the nonlinearity of EEG signals, nonlinear methods, e.g. entropy and fractal dimension analysis, were suggested in the EEG processing of this study. Only one EEG channel of the occipital lobe was selected in the present study, taking into account the importance of real-time applications being fast and the number of channels being small. For fractal size calculation, Higuchi and Box-counting algorithms were employed in the real-time application as feature extraction. In this study, 2D and 3D developed by visual C ++ games were utilized to control the concentration of human subjects. Two sessions were organized to classify the comfortable and concentrated states of EEG signals. Adaptive threshold calculation was used as a classification step. As the concentration of the human subject increased, the game became more entertaining.

Until 2011, only few studies on 2D and 3D have been conducted on the quantitative benefits of 3D visualization systems in comparison to 2D [30]. The change of the P300 response in the oddball paradigm with 2D and 3D stimuli for target-spotting usage was the main framework of the present study. The P300 component was found to be weaker and delayed in 3D stimuli compared to 2D when target identification became more difficult.

In [31], 29 subjects performed virtual spatial navigation and explained the feeling of being there on a 5-point scale. It was determined that the communication between the parietal and frontal lobes of the brain is crucial for a powerful experience of being.

Because of the rapid expansion of 2D and 3D technology fields, understanding brain activation patterns underlying the 2D and 3D comprehensive analyzing, which are to date nearly investigated is considerable [31]. Different studies have been conducted in terms of 2D and 3D entertainment games analysis. The purpose of these 2D and 3D studies was to investigate the degree of eye fatigue [32] when using 2D and 3D game consoles during the game. The regularity of brain signals and how these two conditions make a difference in brain signals were also explored in these studies [8], [33], [4]. In [33], in the five parts of this study, including i) Eyes Closed (EC), ii) Eyes Open (EO), iii) 2D Game Play, iv) 3D Passive Game Play, and v) 3D Active Gameplay, 3D EEG signal complexity was found to be higher than 2D game playing and also from the eyes closed to the eyes open condition. In Mumtaz's study, by using the same dataset of 2D and 3D game playing in five physiological conditions, an acceptable classification success percentage was obtained for these cases.

[4] Explored the response of human factor 2D/3D brain activity in an online modification of the most popular social network, Facebook, and famous game, Angry Birds. The percentage of relative power spectrum analysis for six channels was presented for 2D and 3D games in all EEG bands except for delta. The power spectrum difference of 2D and 3D is shown to be more pronounced in theta and gamma than in other bands.

Besides the game analysis of 2D and 3D technology, its effects on eye fatigue were examined by many researchers [32], [34], [35]. Chen et al. concluded that alpha rhythm significantly decreased in prefrontal and frontal regions after long periods of 3D TV watching. Moreover, a significant decrease was observed in beta activity in prefrontal, frontal, parietal, and temporal regions, and no significant change was found in the mean value of the relative energy in the theta rhythm. Different results in the analysis of fatigue factors in human brain bands were reported by several studies [36], [37], [38].

Based on Brodmann areas, brain areas sensitive to binocular vision and perception of depth were investigated in [39], [40], [41]. Brodmann areas of (BA) BA7 and BA19 of parietal and occipital cortex, respectively, in binocular stereo vision, and BA37 and BA39 areas of the temporal cortex as well as the dorsal regions of the occipital-temporal cortex [39] were sensitive and had an important role in the perception of depth.

In this article, EEG signals of the human brain before and after watching 2D and 3D movies were investigated in detail. The 2D/3D video watching interval was kept short to prevent visual fatigue. As mentioned above, only the power spectrum density (PSD) before and after 2D (I), before and after 3D (II), and after 2D and 3D movie watching (III) stage were calculated in this three-step paradigm. The aim was to see how a power difference occurs in the brain signals before and after watching 2D and 3D movies. By applying PSD based on STFT for all EEG bands of 20 channels, the dominant bands for stages I, II, and III were selected. In Stage III, based on the dominant bands, features were extracted using STFT and DWT to yield a significant difference between R2a and R3a. PLSR and SVM classification algorithms were considered in order to classify the obtained R2a and R3a EEG signals. The main aim of this study was to reveal the power spectrum differences between Relax and Resting steps as a result of 2D and 3D video watching, while the second aim was classifying 2D and 3D Resting steps with a good success percentage by selecting appropriate feature extraction methods. This would enable the creation of criteria for monitoring 2D and 3D TV in an effective and healthy manner.

## 2. Methods and test protocol

### 2.1. Subjects

Five participants with an average age of 31 years took part in this test. The Saw video movie [42] was prepared in 2D and 3D modes for watching. The test and recording protocols were first explained in detail to all participants. All the participants had normal stereo vision, and no candidate had neurological or mental health problems. This study was approved by the Institutional Ethics Committee, with the ethical report number of 24237859-806. Also, all participants signed a written informed consent form before starting the test.

***2.2. Test protocol***

Each signal recording consists of a short relaxation (9 sec.), video watching (14 sec.), and a short rest (9 sec.). Each test was repeated 15 times. A few minutes after the 2D signal recording, exactly the same steps were repeated for the 3D. In general, 75 2D TV and 75 3D TV EEG data sets were collected. According to the LG 32inch, smart TV technology, passive 3D TV glasses were selected.

The EEG signals were acquired by a 21-channel cap (Brain Quick EEG System (Micromed, Italy)). Twenty-one channels Fp1, Fpz, Fp2, F3, F4, F7, F8, C3, C4, Fz, P3, P4, Pz, O1, O2, T3, T4, T5, T6, Oz, and Cz as the reference were placed with the international 10–20 electrode placement system. EEG data were sampled at 512 Hz and the skin impedance was below 10 KΩ. The entire test was performed in a quiet and comfortable environment. All the subjects were asked to avoid unnecessary movements, blinking, nodding, and body movements.

***2.3 Data processing***

To start the band selection step, due to the small amplitude of the EEG signals, values larger than 100 μv were considered as artifacts. In this study, to minimize the effects of noise and artifact on raw EEG data, the average of 15 trials for each subject of 20 channels was calculated. A notch filter of 50 Hz was also applied to suppress the line noise added from the power sources. A third-order zero-phase bandpass Butterworth filter in the frequency range of 1-55 Hz was selected to clean up the noise signal, thus minimizing the non-linear phase effects were minimized. The filter order was experimentally selected as three. To add valuable information to the visual interpretation of the non-stationary and unpredictable nature of EEG data, PSD based on STFT spectrogram was applied in all subdivided EEG bands. Using this time-frequency analysis, a time-varying spectral EEG content image would be presented. This spectrogram is the normalized and the squared magnitude of the coefficients produced by STFT. In large data sets, the use of STFT spectrogram is a logical choice due to its speed and ease of use. Although STFT spectrogram is sufficient for many applications, it has a rough time-frequency resolution due to its window-based processing. In this analysis, it is necessary to pay attention to the width of the window according to the area of use, since a good time resolution and a good frequency resolution cannot be obtained at the same time. Therefore, the relationship between time duration, frequency bandwidth, and the window size is very important.

In this study, the Hanning window with the 512 samples window's length was selected to achieve an acceptable frequency resolution. The overlapping of the window was considered 'windowsize -1'. This window selection with a smoothing characteristic was found to be appropriate because of the different and unpredictable nature of EEG signals.

In this stage, to determine dominant bands for R2b and R2a of Stage I, R3b and R3a of Stage II, and R2a and R3a of Stage III, the following method was applied. The PSD of the EEG signal was calculated; $P_T$ represents total power density in the frequency range of 1 to 49 Hz. This variable was achieved by calculating the area under the curve of the PSD. The MATLAB '*trapz'* command was used for this purpose. $P_\delta$ , $P_\theta, P_\alpha, P_\beta$ and $P_\gamma$ represent the power density at 1-4, 4-8, 8-12, 13-30 and 30-49 Hz, respectively. For each band, the ratio of power to total power was computed. Then, the percentage of normalized power was taken into consideration. These steps were calculated individually for each stage. For the 20 channels, the result was a matrix with the dimensions of 20 * 5 for each mode. The minus of values (R2b-R2a in Stage I, R3b-R3a in Stage II, and R2a-R3a in Stage III) in the matrices of 2D or 3D made the brain-behavior in movie video pre and post-watching. Through these steps and by comparing the results, the dominant bands were determined. Selection of these bands is presented in the Results section. Accordingly, after calculating the normalized power differences in 2D or 3D, depending on the sign, the value of a difference greater than two would be meaningful in the present study. After reviewing the results of all participants, the band of channels meeting this requirement was selected. The gamma band was not taken into account in other sections as it showed the 2D or 3D power spectrum difference in very few channels. In the 2D analysis, δ, θ, α, and β; in 3D, δ, θ, and α; and in Stage III, δ and α were selected and extracted and these bands acted as dominant bands. The flow chart of the band selection is given in Fig. 2 [34].

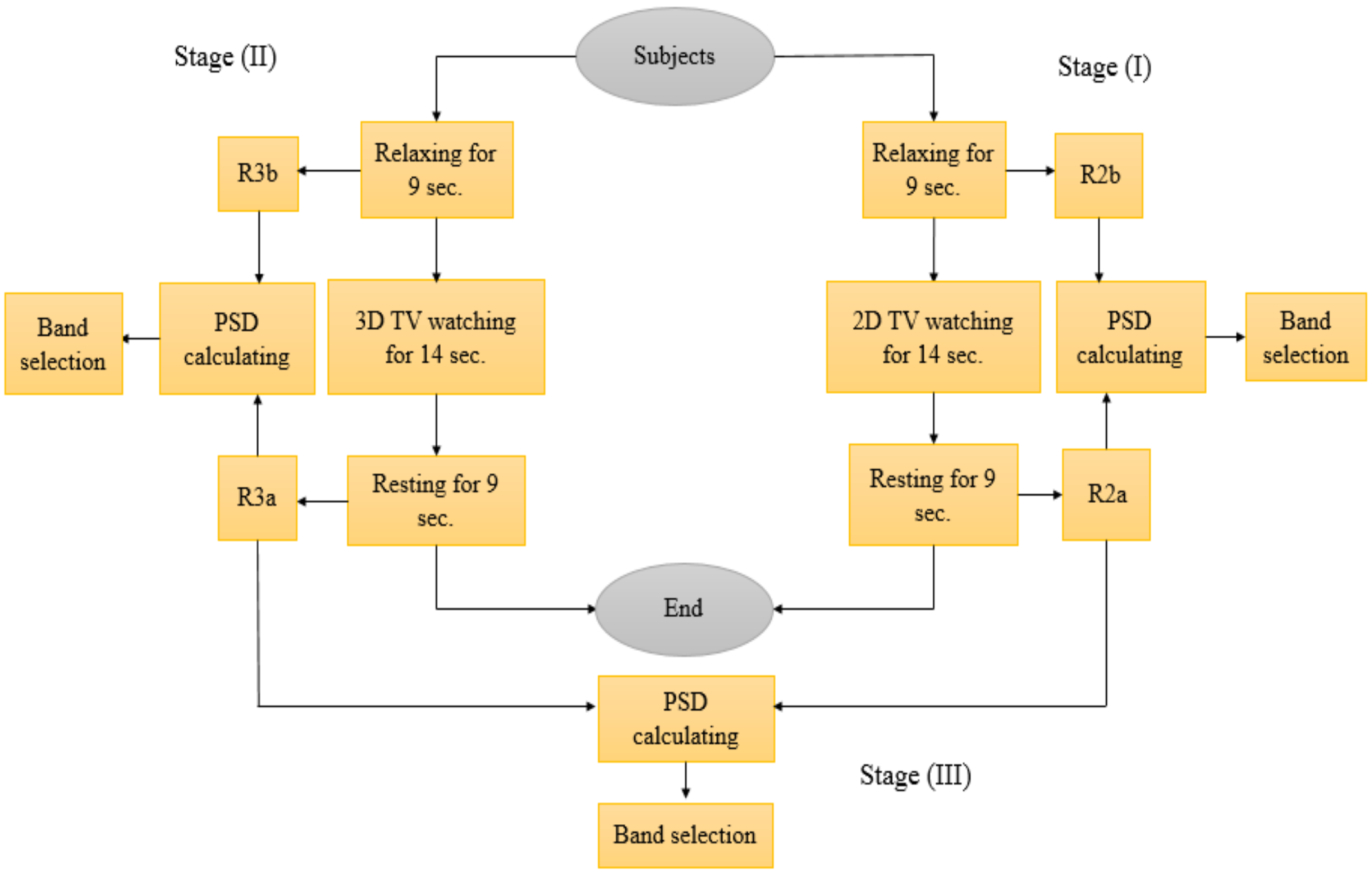

Fig. 2. Flow chart of band selection

### *2.4 Epoch category and feature extraction of Stage III*

After the selection of dominant bands of each stage, only Stage III was taken into account to achieve the other purpose of this article. In this part, the obtained dominant bands of the STFT spectrogram were used to continue the study. In other words, by using stage III's dominant bands, the aim was to extract the effective features of R2a and R3a EEG signals and prepare them for a powerful classification.

In the epoch category, 4sec. time segments with 3.5sec. overlapping were considered. Then, a notch filter at 50 Hz was applied to suppress the line noise. A third-order zero-phase bandpass Butterworth filter in the frequency range of 1-35 Hz was selected to clean up the noise signal. To extract the features, STFT and DWT were performed to the EEG signals of dominant bands for all participants. The Fourier transform in stationary signals is a sufficient factor in the signal analysis, but it is more logical to use time-frequency methods in signals with time-varying and transient characteristics. Thanks to the strong ability of DWT transformation in displaying the time and frequency information of a signal, it is preferred by most researchers [43]. In the STFT feature extraction method, the normalized power of dominant bands was selected as the feature.

By applying DWT, EEG signals were decomposed into approximation and detail coefficients. In order to construct the approximation and detail coefficients of the next level, the approximation coefficients in every level were decomposed once again. Based on the sampling frequency of this study ($F_S$=512 Hz), the EEG epochs were analyzed into various frequency bands using the seventh-order Daubechies (db1) wavelet function up to the 7th level of decomposition. Statistical parameters, including min, max, mean, and standard deviation (SD), were computed for feature extraction [44]. The sub-band decomposition of DWT is presented in Table 2, and Fig. 3. depicts the seven-level wavelet decomposition of the EEG signal.

Table 2. Sub-band decomposition of DWT

| Levels | D1 | D2 | D3 | D4 | D5 | D6 | D7 | A7 |
|---|---|---|---|---|---|---|---|---|
| Frequency ranges (Hz) | 256-512 | 128-256 | 64-128 | 32-64 | 16-32 | 8-16 | 4-8 | 0-4 |

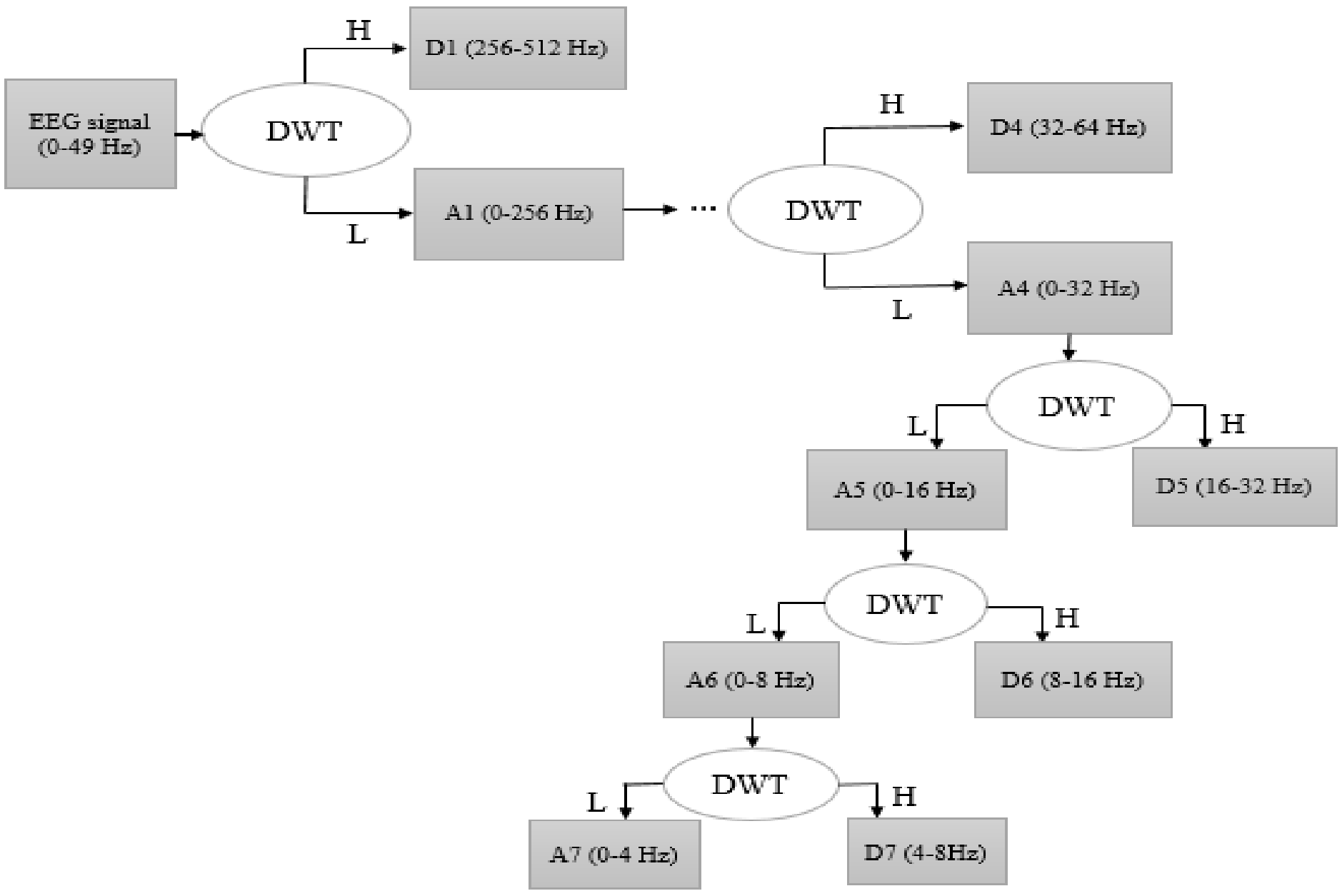

Fig. 3. The seven-level wavelet decomposition of the EEG signal

Dimensions of the feature extraction matrices for 2D/3D would be 4*11*15 and 2*11*15 in DWT and STFT, respectively. For each participant, there were a total of 330½-sec. epochs. There would be two classes in terms of 2D and 3D and 165 epochs per class. In order to prepare a dataset for classification, the epochs of each class were separated into two groups. Next, training and testing sets were prepared. Eighty-three and 82 epochs were chosen for training and testing sets, respectively. The flow chart of the process of data acquisition in Step III is illustrated in Fig. 4 [34].

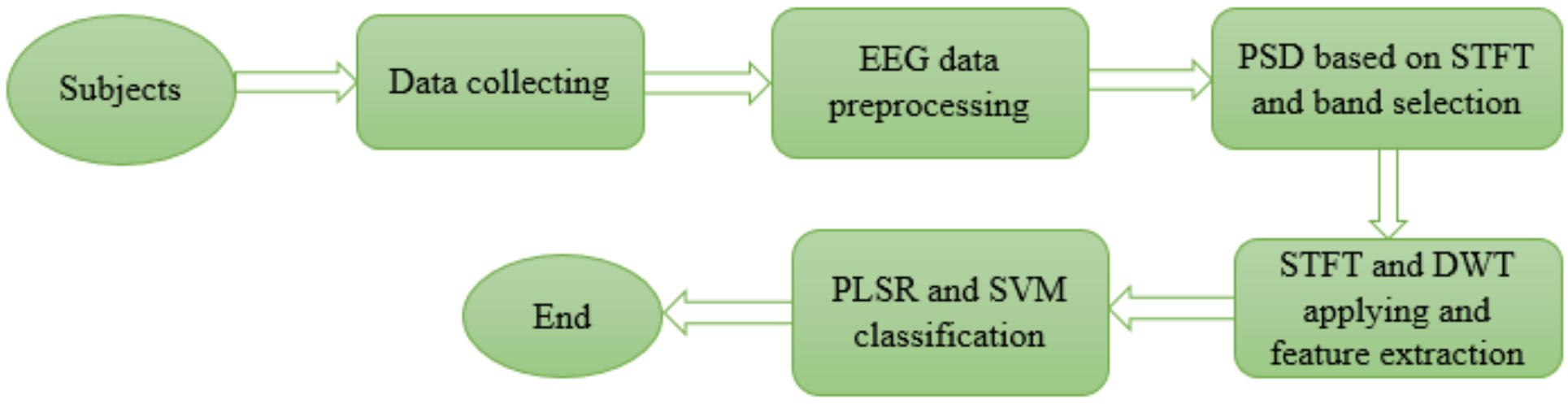

Fig.4. The process of data acquisition and analysis for Step III

### *2.5 PLSR and SVM classification*

In this paper, the classification algorithm consisting of partial least squares regression (PLSR), dimension reduction, and support vector machine (SVM) were compared with one another in diagnosing 2D/3D post-watching states.

To predict the high-dimensional EEG data, PLSR is preferred in the classification context [45], [46], [47]. Besides the statistical discrimination characteristic of PLSR, researchers generally use it as a classifier due to its importance of experimental proof in biomedical fields. Creating a linear model of original variables using latent variables is the main function of this method. The linear combinations of the original variables were gathered in the matrix of features (X) and a vector (y) of class labels [46]. The β distribution of this method has a wide range of applications because of its flexibility. In this paper, to build a set of weight vectors, the nonlinear iterative partial least squares was considered and the basis of this method continued until the desired number of latent vectors were extracted. β is the vector of PLS coefficients. The mathematical expression and details of this method are presented in [48], [49].

It is known in pattern recognition that the classification efficiency is of great importance in obtaining an acceptable performance of research. Generally, the SVM classifier offers good results in problem-solving in high-dimensional, non-linear, and small data sets. The basic principle of the SVM algorithm on input EEG data is finding optimum hyperplanes with a maximum margin between classes by repeating learning iterations [50]. The non-linear kernel function allows the analysis of data to be transferred to the multi-dimensional feature area [51]. It is important to choose a suitable function because of different varieties of the kernel function. Radial basis function (RBF) was used in the present study, and the related kernel function is defined in (1).

$$K(x,y) = exp(\frac{-|x-y|^2}{2\sigma^2}) \qquad (1)$$

In this equation, σ is SD of samples, and x is the feature vector.

### *2.5.1 K-fold cross-validation*

To evaluate the skill of the proposed model, the statistical cross-validation method was employed. By applying this technique, one can learn how the designed model will behave on the new dataset (also known as the test data). In the scientific world, theories are evaluated with predictive performance. In this study, the value of K was taken to be 10. After performing this cross-validation type, the β value of PLSR and the σ value of SVM were obtained.

### *2.5.2 Evaluating the performance of classification*

The defined elements of the confusion matrix in terms of correctly or falsely classified for each class of binary problems are the evaluation measures in classification problems. These elements are defined in Table 3.

Table 3: Confusion matrix

| Predicted value by the test | Actual value as confirmed by the experiment: Positive | Actual value as confirmed by the experiment: Negative |
|---|---|---|
| **Positive** | **TP** | **FP** |
| **Negative** | **FN** | **TN** |

In this table [52], TP are the cases correctly predicted by the classifier in assigning to the positive class, TN are the cases correctly predicted by the classifier in assigning to the negative class, FP are the cases incorrectly predicted by the classifier in assigning to the positive class, and FN are the cases incorrectly predicted by the classifier in assigning to the negative class. In the present study, the 2D class was defined as the positive samples and the 3D class as the negative samples. Based on this table, the performance criteria used in this study, namely accuracy, sensitivity and specificity, are described below.

$$Accuracy = \frac{TP+TN}{TP+TN+FN+FP} \quad (2)$$

$$Sensitivity = \frac{TP}{TP+FN} \quad (3)$$

$$Specificity = \frac{TN}{TN+FP} \quad (4)$$

### 3. *Results*

To investigate the impact of 2D and 3D pre- and post-video watching, PSD based on short-time Fourier transform spectrogram was performed on the processed-EEG data of five participants. As mentioned above in the three stages (I, II, III) for each channel, the average PSD of five EEG bands was calculated to highlight the significant bands of each stage. In this section, we divided the channels representing the scalp into five main groups in order to discuss the results more clearly: frontal (Fp1, Fpz, Fp2, F3, F4, F7, F8, Fz), occipital (O1, O2, Oz), parietal (P3, P4, Pz), central (C3, C4), and temporal (T3, T4, T5, T6). The average difference of PSD for five participants of δ, θ, α, β, and γ bands are is given in Figs. 5 and 6, respectively. It is evident that the γ band does not offer the power difference to be considered in all channels before and after 2D and 3D video watching.

In the δ band of Stage I, the PSD value of R2b was larger than R2a only in meaningful channels of the frontal lobe, i.e. F7, F8, and Fz. This means that the signal's power content was larger before watching than after watching in 2D. In the meaningful channels of the parietal lobe, i.e. P3, P4, and Pz the signal's power content was larger after watching than before watching. For O2 and Oz of the occipital and T6 of the temporal lobe, the PSD value of R2a was larger than R2b.

In the δ band of Stage II, the PSD value of R2a was larger than the before watching state in the Fp2, the situation was vice versa in F7. In meaningful channels of C3, C4, Pz, O1, O2, Oz, T3, T5, and T6, the PSD value of R2b was larger than R2a.

In the θ band of Fig. 5, except for F8, all the channels of the frontal lobe and also C3 and C4 of central lobe were channels showing a higher PSD before watching. In O1, Oz, T3, T4, and T5, the PSD value of R2a was larger than R2b.

In the θ band of Stage II, Fp2, F3, F4, C3, P3, P4, and Pz were channels demonstrating a higher PSD before watching, and the situation was vice versa in F7, Fz, O1, and T4.

In the α band of Stage I, in all channels of the frontal lobe, the signal's power content was larger after watching than before watching. The same applied to the C4 and T4 channels. T3, T5, T6, O1, and Oz are channels indicating a higher PSD before watching.

In the α band of Fig. 6. for all the meaningful channels, i.e. F4, F7, C3, P3, Pz, O1, O2, T3, T4, T5, T6, and Oz, the signal's power content was larger after watching than before watching.

In the β band, in all channels of frontal and central lobes except for Fz, Fp2, and C3, the PSD value of R2a was larger than R2b. In Fz, P4, Pz, O1, Oz, T4, and T6, the situation was vice versa.

In the β band of Stage II, T4 was the channel indicating a higher PSD before watching, and P4, Pz, O2, and T5 were channels showing a higher PSD after watching.

According to these results and the PSD difference value variations in many channels of 2D and 3D pre- and post-watching, the dominant bands were selected as δ, θ, α, and β in Stage I, and δ, θ, and α in Stage II.

Stage III was considered to reveal what happens in human brainwaves after watching 2D and 3D videos. At this stage, the signals from the human brain were compared with one another using the effective feature extraction method and classification algorithms. PSD differences were observed in more channels in δ and α bands. Thus, at this stage, these bands were chosen as the dominant bands for the continuation of the study. The average difference of PSD (R2a)-PSD (R3a) for five participants of δ, θ, α, and β bands is depicted in Fig. 7. The gamma band was not taken into account since there was no difference in the PSD of many channels after watching 2D and 3D videos.

In the δ band of this stage, in almost all of the meaningful channels except for T4, the PSD of 2D viewers had a positive value when compared to the 3D one, and at T4, the condition was vice versa. In the α band, all the meaningful channels, the PSD of the 3D viewer was more positive than the 2D one. At T5, for 2D and 3D post-video viewers, the maximum difference of the average PSD was observed. Moreover, the P3 channel of the band had the maximum PSD difference among 20 channels in 2D and 3D after watching the video.

As mentioned above, the first aim of this study was to investigate the PSD of human brainwave before and after watching 2D/3D videos at all EEG channels of five frequency bands. In this PSD analysis, the meaningfully affected channels and their related EEG frequency bands were determined. These bands were those separately representing Stages I, II and III as the dominant band. The meaningful and effective channels of each stage were separately included in the analysis. The behavior of these channels is shown in Figs. 5, 6, and 7 for Stages I, II, and III.

After the dominant band selection, the second objective of this study was to classify the two Resting stages after watching 2D and 3D videos. Two methods were followed for classifications. First, STFT was utilized after the dominant band selection, and in the second method, DWT was used as feature extraction. The average of PLSR and SVM classification results using STFT and DWT for five subjects is shown in Figs. 8 and 9 respectively.

By a general overview of these classification results, one can conclude that the SVM classifier and DWT feature extraction method yield a better result compared to PLSR and STFT. As this study is a two-class study, we have used the best combination of channels to bring the results from the classifiers to an acceptable level. In order to improve the results of the proposed classifiers, for each classification method, both PLSR and SVM algorithms for these combinations were tested. These channels and their combinations were tested with PLSR and SVM classifiers based on STFT and DWT feature extraction and are shown in Figs. 10, 11, 12, and 13. These selected channel combinations play an important role in increasing classifier results. Furthermore, in STFT, the classification percentage of success was observed up to 73.36% and 89.11% increase for PLSR and SVM, respectively. For DWT, these results were 75.18% and 87.26%. Besides the classification accuracy, the average sensitivity and specificity of all the participants for SVM classifier and DWT feature extraction were calculated. These performance parameters had good compromises with one another in this study. In the best channel combinations of SVM classifier, when the best channels were ranked, the success rate of PLSR and SVM was shown to increase to 82% and 97% in STFT and to 83% and 95% in DWT.

In addition to improving the classifier result, it is important to achieve good success in biomedical studies using fewer EEG channels. Consequently, the compromise between the high success percentage and fewer number of channels is quite important. Accordingly, since the SVM classification has too many channels, the best channel combination of O2, F7, T4, T6, F3, T3, F8 and T5 for STFT and F8, Fpz, Fp2, T4, F7, and Fp1 for DWT were selected based on the acceptable classification result. When looking at these channel combinations, it was observed that occipital, temporal, and frontal lobes were suitable lobes for classifying the Resting process after watching 2D/3D movies.

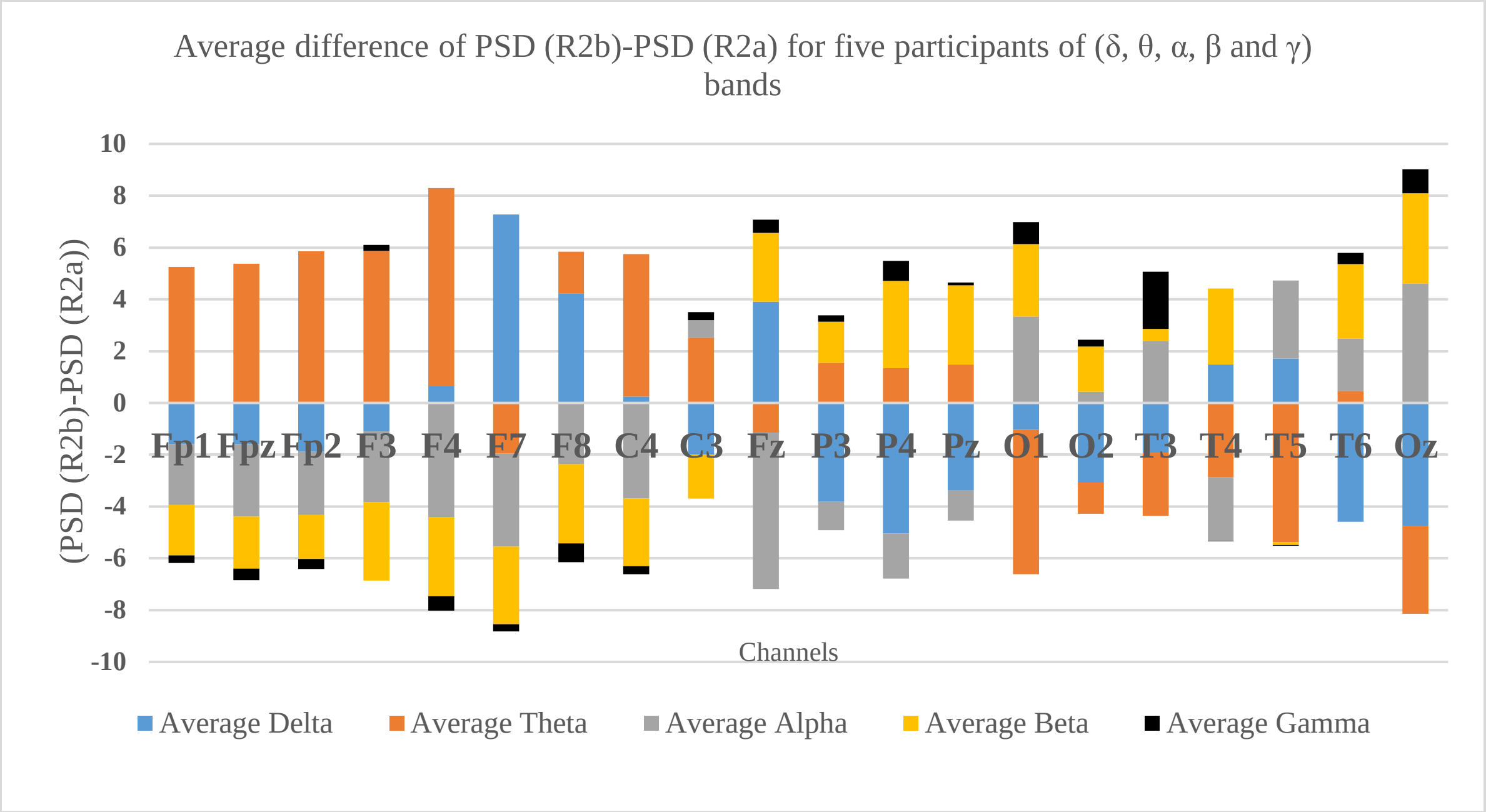

Fig. 5. Average difference of PSD (R2b)-PSD (R2a) for five participants of the δ, θ, α, β and γ bands

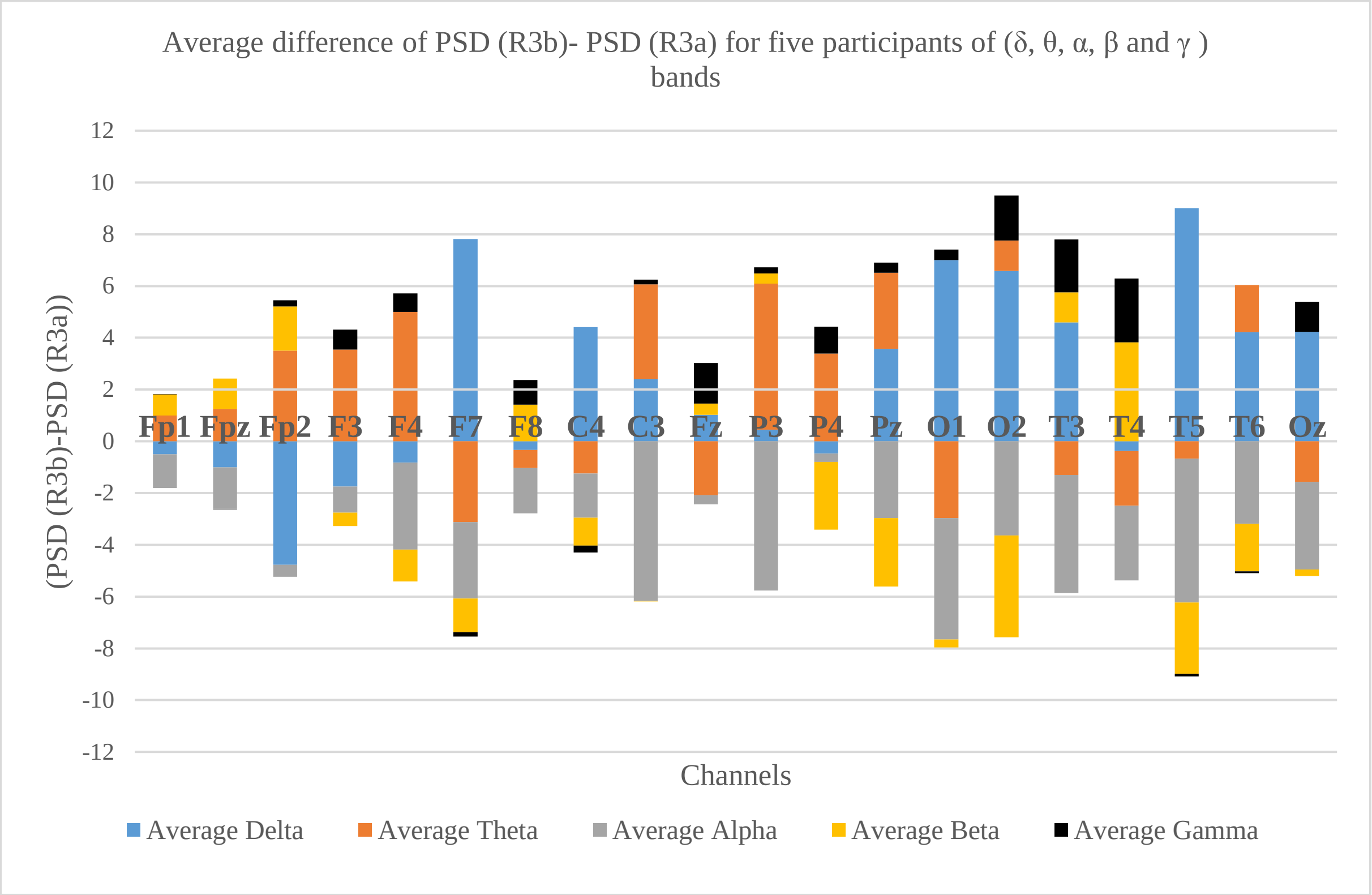

Fig. 6. Average difference of PSD (R3b)-PSD (R3a) for five participants of δ, θ, α, β and γ bands

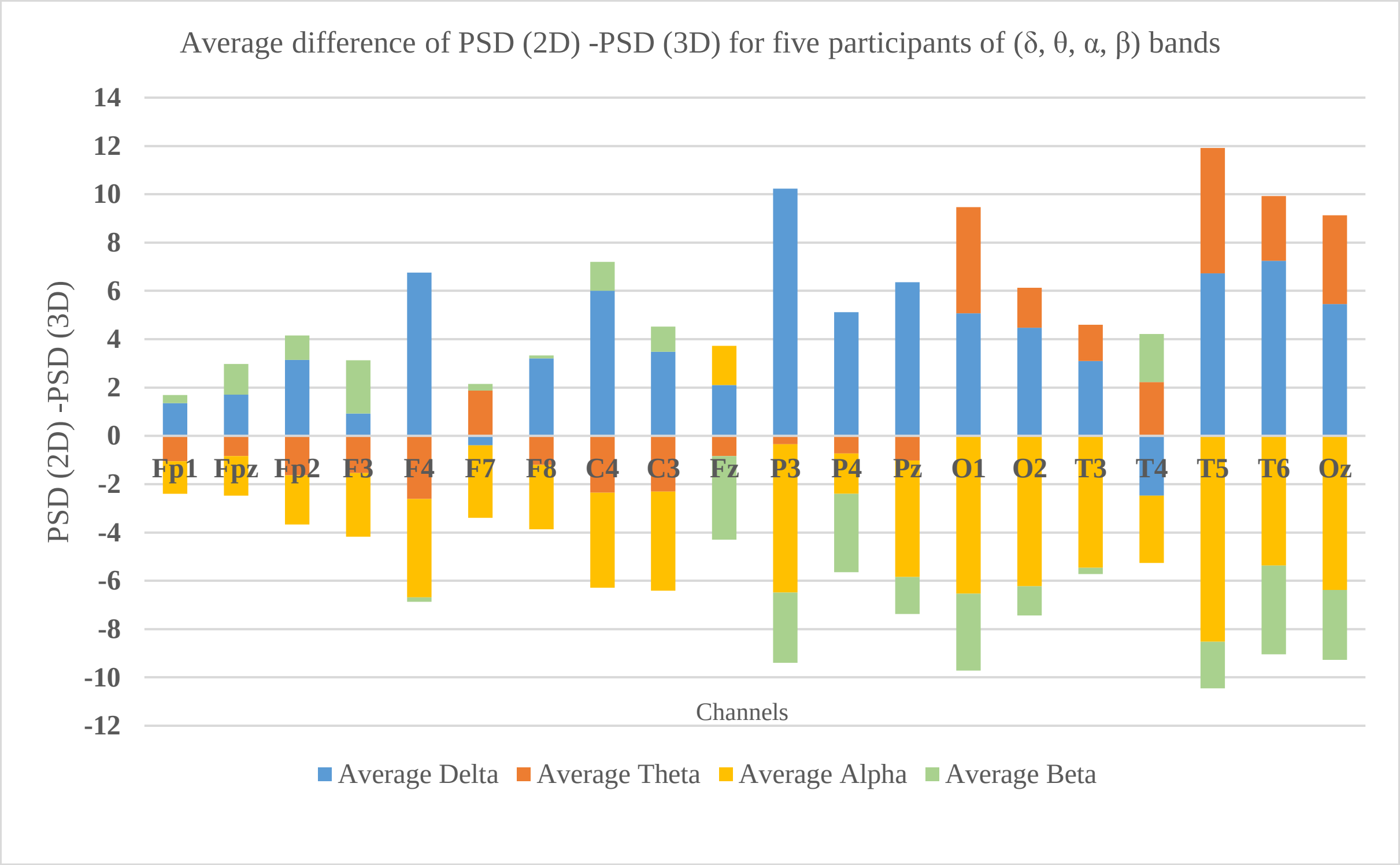

Fig. 7. Average difference of PSD (R2a)-PSD (R3a) for five participants of δ, θ, α, and β bands

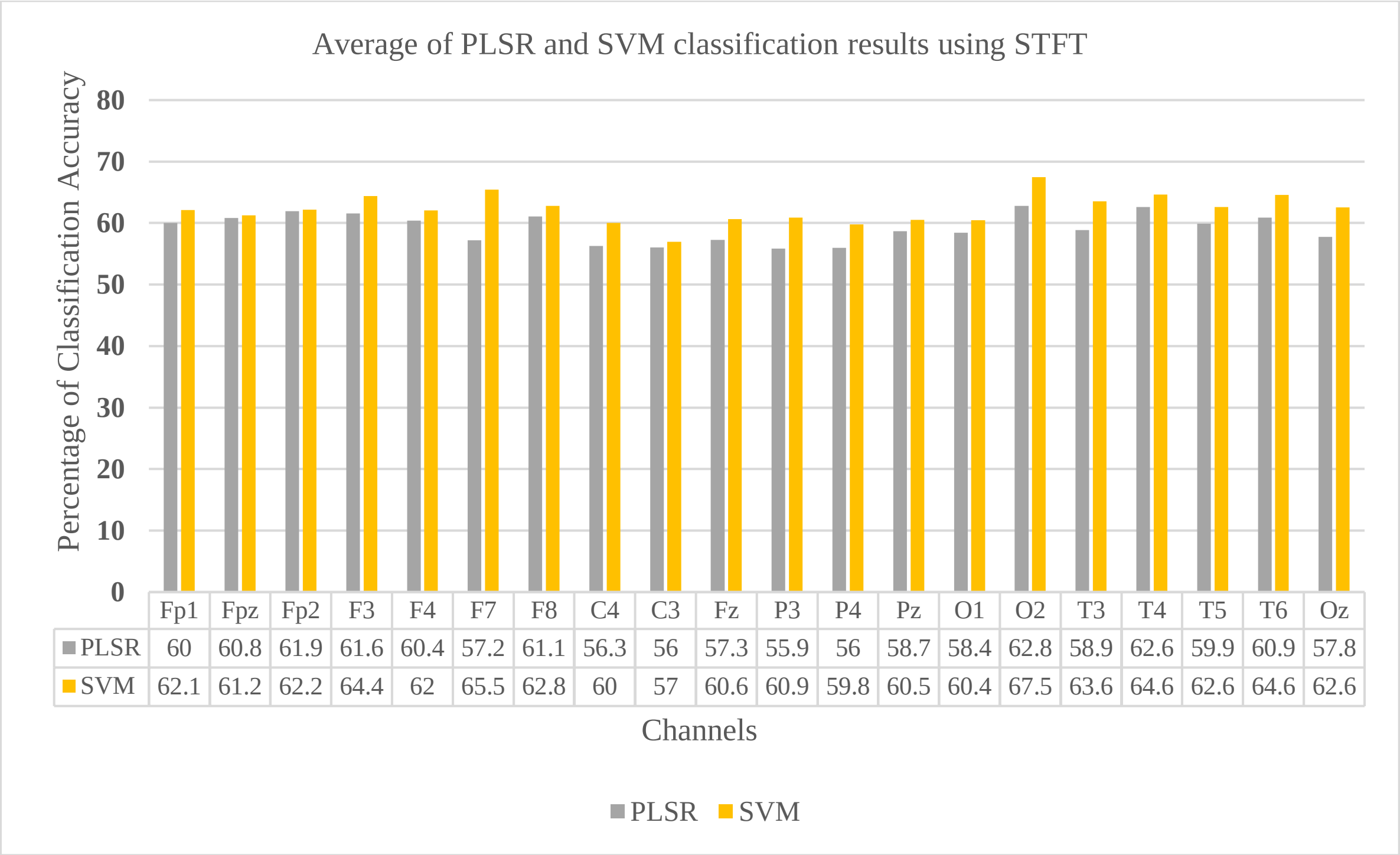

Fig. 8. Average of PLSR and SVM classification results using STFT

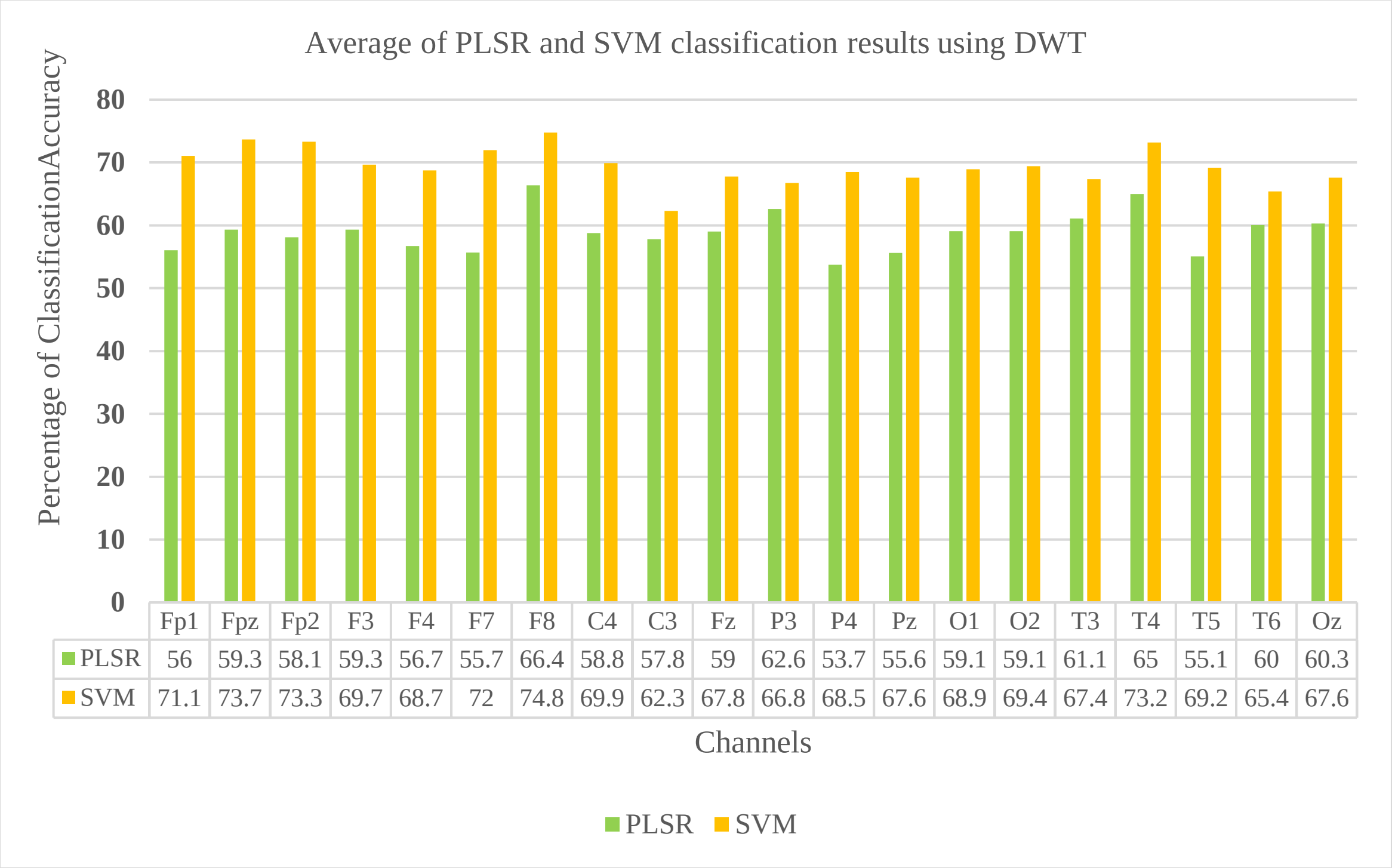

| | Fp1 | Fpz | Fp2 | F3 | F4 | F7 | F8 | C4 | C3 | Fz | P3 | P4 | Pz | O1 | O2 | T3 | T4 | T5 | T6 | Oz |
|---|---|---|---|---|---|---|---|---|---|---|---|---|---|---|---|---|---|---|---|---|
| PLSR | 56 | 59.3 | 58.1 | 59.3 | 56.7 | 55.7 | 66.4 | 58.8 | 57.8 | 59 | 62.6 | 53.7 | 55.6 | 59.1 | 59.1 | 61.1 | 65 | 55.1 | 60 | 60.3 |
| SVM | 71.1 | 73.7 | 73.3 | 69.7 | 68.7 | 72 | 74.8 | 69.9 | 62.3 | 67.8 | 66.8 | 68.5 | 67.6 | 68.9 | 69.4 | 67.4 | 73.2 | 69.2 | 65.4 | 67.6 |

Fig. 9. Average of PLSR and SVM classification results using DWT

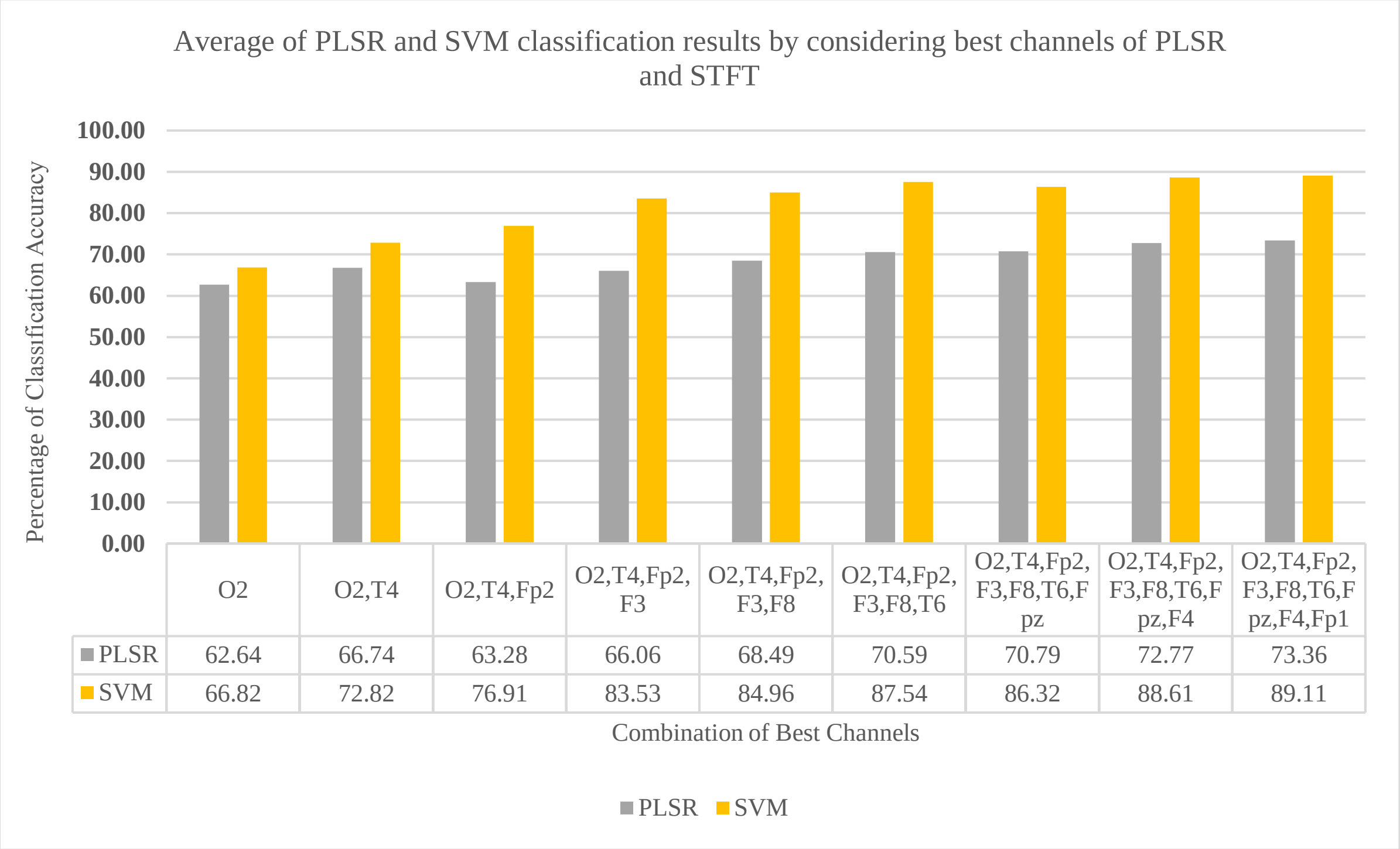

| | O2 | O2,T4 | O2,T4,Fp2 | O2,T4,Fp2, F3 | O2,T4,Fp2, F3,F8 | O2,T4,Fp2, F3,F8,T6 | O2,T4,Fp2, F3,F8,T6,F pz | O2,T4,Fp2, F3,F8,T6,F pz,F4 | O2,T4,Fp2, F3,F8,T6,F pz,F4,Fp1 |
|---|---|---|---|---|---|---|---|---|---|
| PLSR | 62.64 | 66.74 | 63.28 | 66.06 | 68.49 | 70.59 | 70.79 | 72.77 | 73.36 |
| SVM | 66.82 | 72.82 | 76.91 | 83.53 | 84.96 | 87.54 | 86.32 | 88.61 | 89.11 |

Fig. 10 Average of PLSR and SVM classification results by considering the best channels of PLSR and STFT

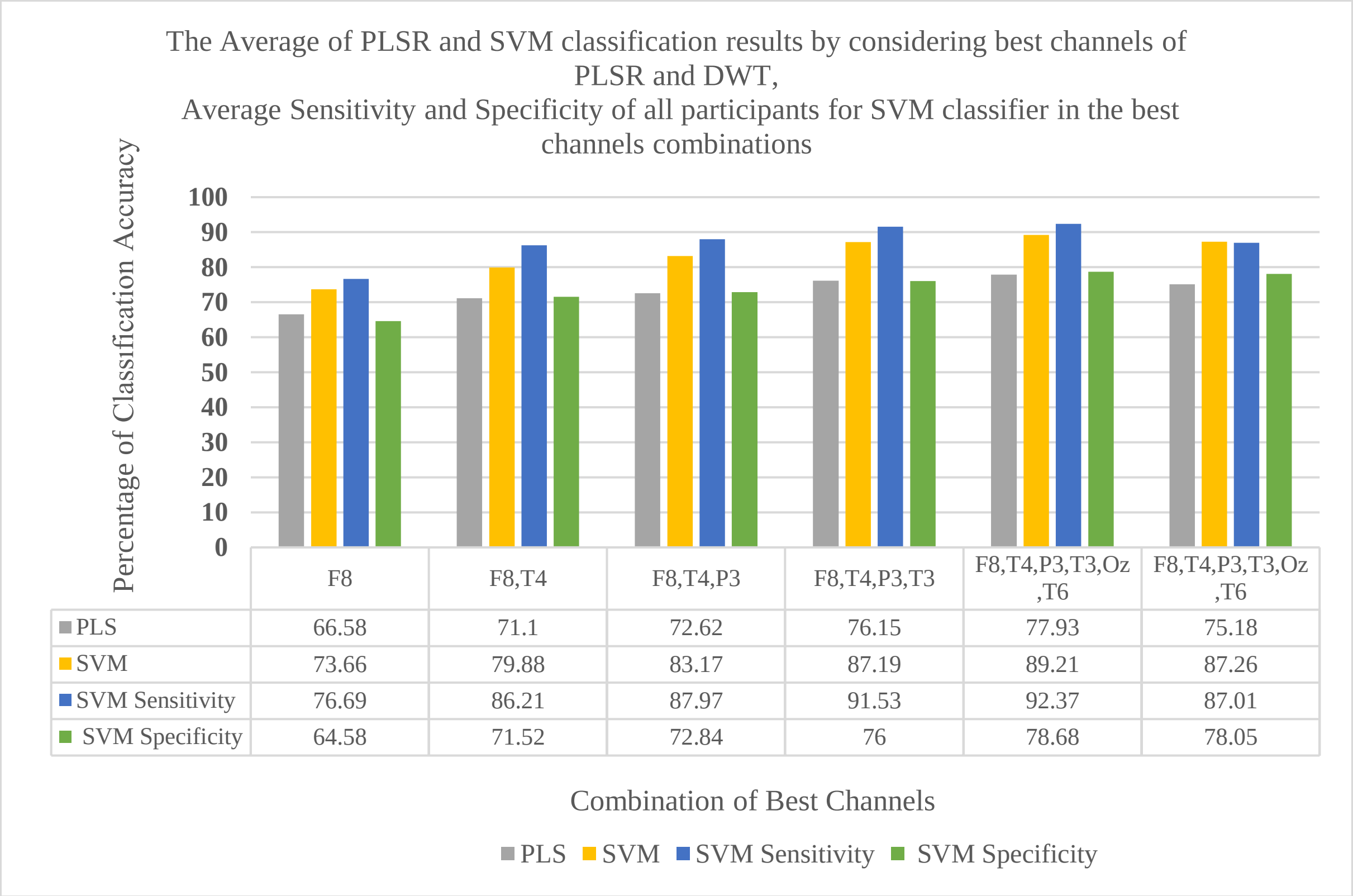

Fig. 11. Average of PLSR, SVM classification results, sensitivity, and specificity by considering the best channels of PLSR and DWT

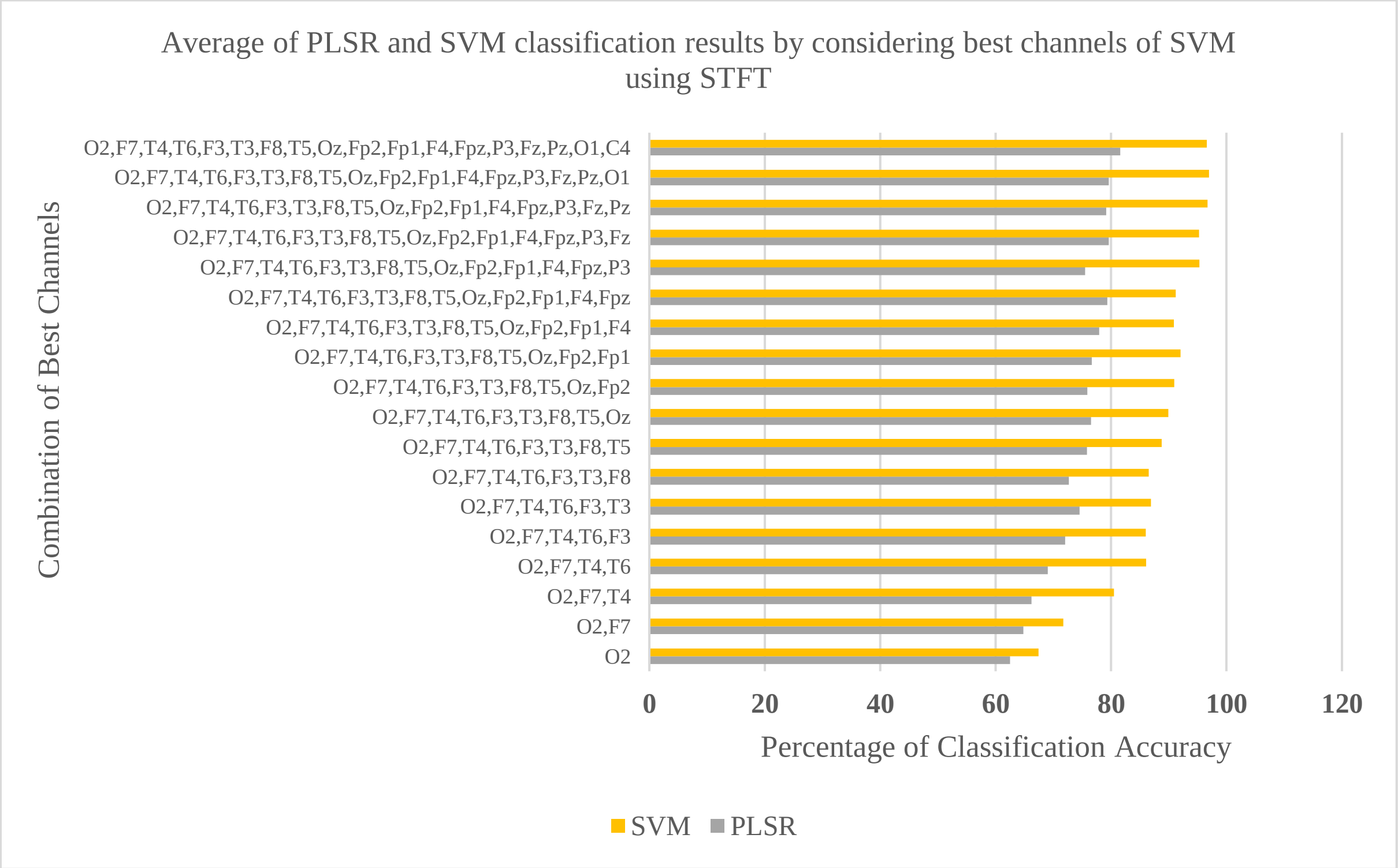

Fig. 12. Average of PLSR and SVM classification results by considering the best channels of SVM using STFT

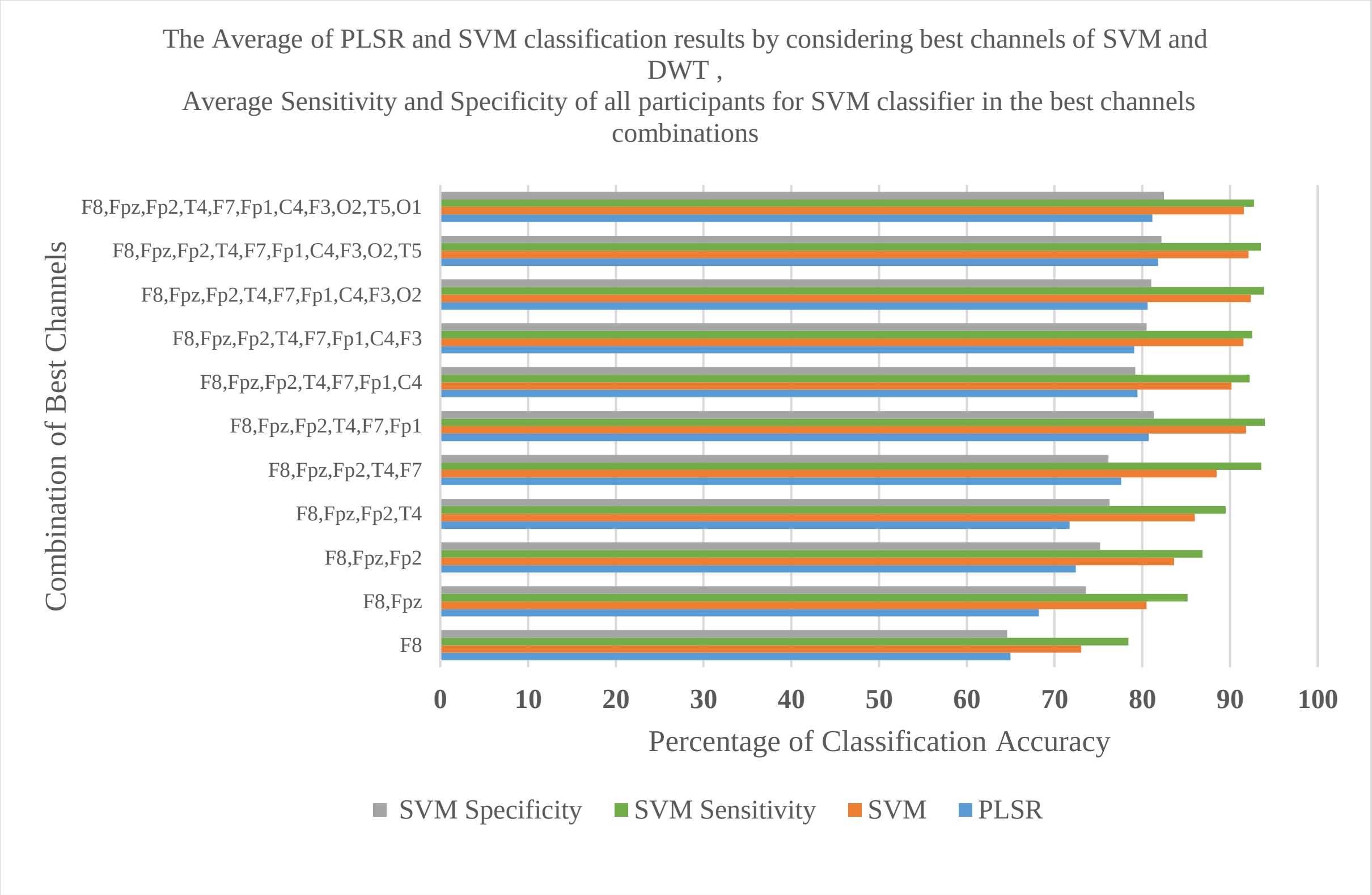

Fig. 13. Average of PLSR, SVM classification results, sensitivity, and specificity by considering the best channels of SVM using DWT

### *4. Discussion*

To date, the effects of 2D/3D pre- and post-movie watching have not been well established. Thus, a detailed quantitative study was performed to fill this gap. When this study is expanded, it can shed light on factors affecting eye fatigue and also human brain responses predicting the 2D/3D video watching. The most important feature of this study is that it took into consideration the behaviors of all bands and EEG channels throughout the research. In nearly all EEG studies of 2D/3D, based on some basic information, only some bands and channels were analyzed by default [53], [54], [39].

In this research, the PSD of the brain signals of 2D/3D pre and post-video movie viewers in all EEG bands were explored, and then the dominant bands of each stage were selected. In 2D/3D studies, there is no clear decision about the behavior of the brain waves in EEG bands. The majority of studies in this domain have presented different ideas and results. In this study, for each channel of each band, the dominant bands which included more channels with maximum PSD difference were selected. The magnitude of this value in meaningful channels may represent eye strain after watching a film. These changes in the PSD of EEG channels show that 3D TVs cause more visual fatigue compared to 2D [28].

As expected, in feature extraction, the DWT method had superior results than STFT [55]. The average PLSR and SVM classification results for each channel made the meaningful channels of this study more significant. It is worth mentioning that the channels representing regions sensitive to stereo vision and depth perception were included in our selected channels that gave the best results in this study. Channels representing the back part of the cortex or the occipital lobe were important in this study thanks to their ability in interesting visual information. These channels were included in the best channel combinations. The best channel combinations also contained channels representing the temporal lobe [56]. In addition to their ability in understanding speech, these channels are thought to have finer settings to represent visual objects in the middle parts of the temporal lobe [57]. Channels representing the frontal lobe were also included in the best channel combinations. The responsibilities of this lobe include speech, some motor skills, attention management, emotion control, and the evaluation of similarities and differences between the two objects.

***5. Conclusion***

This paper dealt with a novel method of EEG signals analysis using band selection and the best channel combination. After a suitable feature extraction method selection, PLSR and SVM classification algorithms were applied to examine the effects of 2D/3D pre- and post-video watching. The PSD of the EEG recording of 2D/3D film viewers was selected as the key parameter of this study. The analysis showed significant differences between 2D/3D pre- and post-video watching. In the band selection of Stage I, all EEG bands except the gamma were seen as the dominant and effective bands in Relaxation and Resting. In Stage II, delta, theta, and alpha were the significant bands. Among various EEG frequency bands, delta, and alpha were selected for analysis in Stage III due to their relationship with the relaxation of the brain. Based on the results of this study, in dominant bands, it was demonstrated that the temporal, occipital, and frontal lobes were more effective in classifying the stages after watching 2D/3D videos.

In future research, the number of participants can be increased to solve one of the limitations of this study. In this study consisting of a three-stage paradigm, it was aimed to analyze the moment of transition from Watching the video to Relax and from Watching to Rest. Our future study will aim to reduce the number of channels and increase the classification accuracy by testing different feature extraction and classification methods. Different power ratios can also be used in the 2D/3D analysis.